\documentstyle[12pt,epsf]{article}

\begin{document}

\begin{center}
{\bf  SPIN ASYMMETRIES IN DIFFRACTIVE $J/\Psi$ LEPTOPRODUCTION.}

\vskip 5mm
S.V.Goloskokov$^{\dag}$

\vskip 5mm

{\small
{\it
Bogoliubov Laboratory of Theoretical  Physics,\\
 Joint Institute for Nuclear Research,\\
Dubna 141980, Moscow region, Russia
}
\\
$\dag$ {\it
E-mail: goloskkv@thsun1.jinr.ru
}}
\end{center}

\vskip 5mm

\begin{center}
\begin{minipage}{150mm}
\centerline{\bf Abstract}
In this report we calculate the cross section and $A_{ll}$
asymmetry for the diffractive $J/\Psi$ leptoproduction.
We  study dependences of the asymmetry on the
structure of the Pomeron-proton coupling.
\end{minipage}
\end{center}
\vskip 10mm

The study of the Pomeron nature becomes again popular now due
to the progress in investigation of diffractive reactions at HERA
\cite{h1_zeus}. An important information on the nonperturbative
Pomeron structure can be obtained from the processes of
diffractive vector meson leptoproduction. Different models have
been proposed to study the $\gamma^* p \to V P$ reaction.  The
QCD-inspired models based on the two-gluon form of the Pomeron
have been used in \cite{rys,mod} to analyze the $Q^2, t$ and energy
dependences of elastic vector meson production. In papers
\cite{rys} it has been found  that the cross-section of the
vector-meson photoproduction is proportional to $[x
G(x)]^2$ at high $Q^2$. The sensitivity of diffractive lepto-- and
photoproduction to the gluon density in the proton can give an
excellent tool to test G(x) in these reactions. The connection of
spin asymmetry in diffractive reactions with the gluon distribution
$\Delta G$ is not well known. Thus, it is very important to
perform the model calculations of spin-dependent $J/\Psi$
leptoproduction to obtain the numerical estimations of spin
asymmetries.

In this report we shall calculate the cross section and the

$A_{ll}$ asymmetry of the diffractive $J/\Psi$ leptoproduction
\begin{equation}
l+p \to l+p +J/\Psi \label{react}
\end{equation}
at high energies and small $Q^2$. We shall study the sensitivity
of the asymmetry to the spin structure of the Pomeron-proton
coupling.

The cross section of $J/\Psi$ production can be
decomposed into the leptonic and hadronic tensors, the amplitude
of the $\gamma^\star I\hspace{-1.7mm}P \to J/\Psi$ transition and
the Pomeron exchange.
In the model \cite{lansh-m}, the Pomeron effectively couples to
the hadron like a $C= +1$ isoscalar photon. Then, the Pomeron
coupling with the proton is equivalent to the isoscalar
electromagnetic  nucleon current with the Dirac and Pauli nucleon
form factors \cite{nach}. It can be written in the form
\begin{equation}
\label{ver}
V_{pgg}^{\alpha}(p,t,x_P)= 2 p^{\alpha}
A(t,x_P)+\gamma^{\alpha}  B(t,x_P),
\end{equation}
where $r$ is the momentum transfer the proton experiences and
$x_P$ is a fraction of the initial proton momentum carried by the
gluon system. The term proportional to $B(r)$ represents the
standard Pomeron coupling that leads to the non-flip amplitude.
The $A(r)$ amplitude  produces the spin-flip effects in the
Pomeron coupling which do not vanish at high-energies. A similar
form of the proton-Pomeron coupling has been used in \cite{klen}
to analyze the spin effects in diffractive vector-meson
production. In model \cite{gol_mod},  form (\ref{ver}) was
found to be valid for small momenta transfer $|t| <  \mbox{few
GeV}^2$ and the $A(t)$ amplitude to be caused  by the meson-cloud
effects in the nucleon. In a QCD--based diquark model of the
proton \cite{kroll},  structure (\ref{ver}) of the $pgg$
coupling  has been found for moderate momenta transfer
\cite{gol_kr}. The $A(t)$ contribution there is determined by the
effects of vector diquarks inside the proton, which are of an
order of $\alpha_s$.

In our model we shall use the simple form (\ref{ver}) of the
Pomeron-proton coupling. This will allow us to calculate the
spin-dependent cross-section. The gluons from the Pomeron are
coupled with  single and different quarks in the $c \bar c$
loop. This ensures the gauge invariance of the final result
\cite{diehl}. In what follows we shall consider the $J/\Psi$
meson as an $S$-wave system of $c \bar c$ quarks \cite{berger}. The
$J/\Psi$-wave function then has a form $g (\hat k+m_c)
\gamma_\mu$ where $k$ is the momentum of $c \bar c$- quarks. In
the nonrelativistic approximation the mass of the $c$ quark should be
equal to $m_J/2$. The coupling constant $g$ can be expressed
through the $e^+ e^-$ decay width of the $J/\Psi$ meson
\begin{equation}
g^2=\frac{3 \Gamma^J_{e^+ e^-} m_J}{64 \pi \alpha^2}.
\end{equation}

The hadronic tensor
for vertex (\ref{ver}) has the form
\begin{equation}
\label{wtenz}
W^{\alpha,\alpha'}(s^p)= \sum_{final spin} \bar u(p')
V_{pgg}^{\alpha}(p,t,x_P) u(p,s^p)
\bar u(p,s^p)
V_{pgg}^{\star\,\alpha'}(p,t,x_P) u(p').
\end{equation}
Here $p$ and $p'$ are  initial and final proton momenta,
and $s^p$ is a spin of the initial proton.
The sum and difference of the cross sections with parallel and antiparallel
longitudinal polarization of a lepton and a proton are expressed in terms of the
spin-average and spin--dependent value of the lepton and hadron
tensors which are determined by the relation

\begin{equation}
W^{\alpha,\alpha}(\pm)=\frac{1}{2}( W^{\alpha,\alpha'}(+\frac{1}{2})
\pm W^{\alpha,\alpha'}(-\frac{1}{2})),
\end{equation}
where $W(\pm\frac{1}{2})$ are the tensors with the
helicity of the initial proton or lepton equal to $\pm 1/2$.

The leading term of the spin average hadron tensor looks like
\begin{equation}
W^{\alpha;\alpha'}(+) = 4 p^{\alpha} p^{\alpha'}  [ |B(t)+2 m A(t)|^2 + |t| |A(t)|^2].
\end{equation}
It is proportional to the proton-proton cross section up to a
 function of $t$.
The  spin-dependent hadron tensor is quite complicated:
\begin{eqnarray}
W^{\alpha;\alpha'}(-) = 2 i [ |B(t)|^2  \epsilon^{\alpha\alpha'\delta\rho} (p'-p)_{\delta}
s^p_{\rho}
-2/m A^\star(t) B(t) p^{\alpha'}\epsilon^{\alpha\gamma\delta\rho}
p_{\gamma} (p'-p)_{\delta} s^p_{\rho} \nonumber\\
 + 2/m A(t) B^\star(t) p^{\alpha} \epsilon^{\alpha'\gamma\delta\rho}
p_{\gamma} (p'-p)_{\delta} s^p_{\rho}].
\end{eqnarray}
The first term here contains  indices of
different Pomeron couplings in the $\epsilon$ function. This
contribution is equivalent to the spin-dependent part in the lepton
tensor, and we call it
a full block asymmetry. The other terms are products of the asymmetric
part of one--proton vertex by the symmetric part of the other.

The cross section of $J/\Psi$ leptoproduction can be written in
the form
\begin{equation}
\frac{d\sigma^{\pm}}{dQ^2 dy dt}=\frac{|T^{\pm}|^2}{32 (2\pi)^3
 Q^2 s^2 y}, \label{ds}
\end{equation}
For the spin-average  amplitude square we find
\begin{equation}
 |T^{+}|^2=  N ((2-2 y+y^2) m_J^2 + 2(1 -y) Q^2) s^2 [|B+2 m A|^2+|A|^2 |t|] I^2.
\label{t+}
\end{equation}
Here $N$ is a known normalization factor and $I$ is the integral
 over transverse momentum of the gluon
\begin{eqnarray}
I=\frac{1}{(m_J^2+Q^2+|t|)}
\int \frac{d^2l_\perp (l_\perp^2+\vec l_\perp \vec \Delta)}
{(l_\perp^2+\lambda^2)((\vec l_\perp+\vec \Delta)^2+\lambda^2)[l_\perp^2+
\vec l_\perp \vec \Delta
+(m_J^2+Q^2+|t|)/4]}.
\end{eqnarray}
The term proportional to $(2-2 y+y^2) m_J^2$ in (\ref{t+})
represents the contribution of a virtual photon  with transverse
polarization. The $2(1 -y) Q^2$ term describes the effect of
longitudinal photons.

We shall integrate the cross sections (\ref{ds}) over $Q^2$ and
$y$
\begin{equation}
\frac{d\sigma^{\pm}}{dt}=\int_{y_{min}}^{y_{max}} dy
\int_{Q^2_{min}}^{Q^2_{max}} dQ^2
\frac{d\sigma^{\pm}}{dQ^2 dy dt},
\end{equation}
where
\begin{equation}
Q^2_{min}=m_e^2\frac{y^2}{1-y};\;\; Q^2_{max} \sim 4 \mbox{GeV}^2
\end{equation}
The cross section for $J/\Psi$ production at HERA energy
$\sqrt{s}=300GeV^2$
is shown in Fig.1.\\

\begin{minipage}{7.3cm}
\vspace{.1cm}
\hspace{-.7cm}
\epsfxsize=7.1cm
\epsfbox{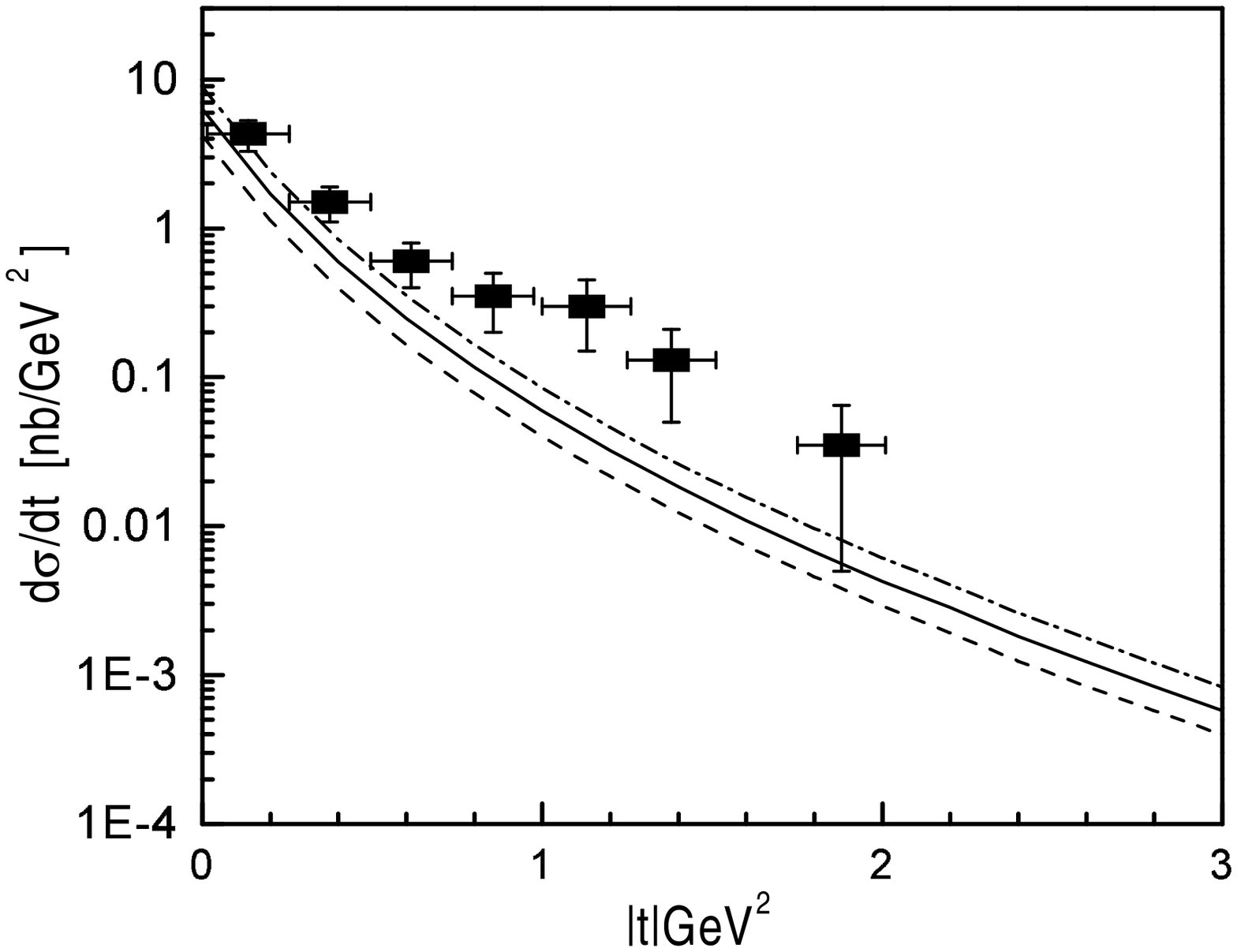}
\end{minipage}
\begin{minipage}{0.cm}
\end{minipage}
\vspace{-.1cm}
\begin{minipage}{7.3cm}
\epsfxsize=7.0cm
\epsfbox{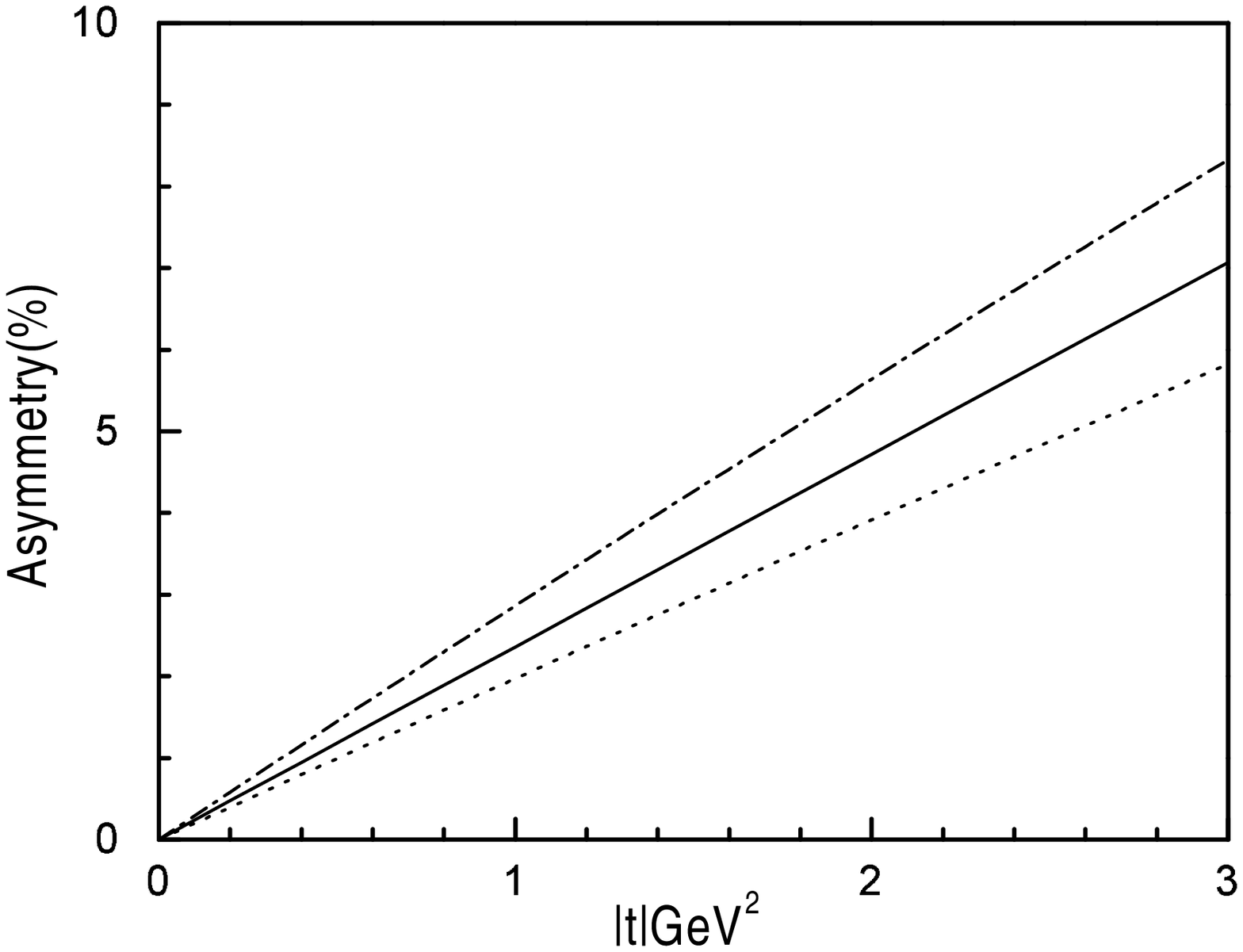}
\end{minipage}
\vspace{.3cm}
\phantom{The cross section for $J/\Psi$ production at HERMES energy$s=50GeV^2$are
shown in Fig.1.}
\begin{minipage}{7.3cm}
Fig.1~ Cross section for $J/\Psi$ production at HERA energies:
solid line -for $\alpha_{flip}=0$; dot-dashed line -for
$\alpha_{flip}=-0.1$;
dashed line -for $\alpha_{flip}=0.1$.
\end{minipage}
\begin{minipage}{.1cm}
\phantom{aaa}
\end{minipage}
\begin{minipage}{7.2cm}
Fig.2~ $A_{ll}$ asymmetry of $J/\Psi$ production  at HERMES:
solid line -for $\alpha_{flip}=0$; dot-dashed line -for
$\alpha_{flip}=-0.1$;
dashed line -for $\alpha_{flip}=0.1$.
\end{minipage}
\phantom{The cross section for $J/\Psi$ production at HERMES energy $s=50GeV^2$are
shown in Fig.1.}
The spin-dependent amplitude square looks as
\begin{equation}
 |T^{-}|^2= N (2- y)  s |t|  [|B|^2+ m (A^\star B +A B^\star)] m_J^2 I^2.
\label{t-}
\end{equation}
As a result, we find the following form of asymmetry
\begin{equation}
\label{asy}
A_{ll} \sim \frac{|t|}{s}\frac{[|B(t)|^2+ m (A(t)^\star B(t)+ A(t) B(t)^\star)]}
{(2-2 y+ y^2)[|B(t)+2 m A(t)|^2+ |t| |A(t)|^2]}.
\end{equation}

The predicted asymmetry at the HERMES energy as a function of the
ratio of spin-flip to non-flip  parts of the Pomeron coupling
$\alpha_{flip}=A(t)/B(t)$ is shown in Fig.\ 2. The important
property of  $A_{ll}$  is that the asymmetry of vector meson
production is equal to zero for the forward direction. The
$A_{ll}$ asymmetry might be connected with the spin-dependent
gluon distribution $\Delta G$ only for $|t|=0$. Thus, $\Delta G$
cannot be extracted from $A_{ll}$ in agreement with the results
of  \cite{mank}. The rapid energy dependence of asymmetry is
another important property of (\ref{asy}). It has been shown in
\cite{gola_ll} that the $A_{ll}$ asymmetry in the diffractive
processes is proportional to the fraction of the initial proton
momentum $x_p$ carried off by the Pomeron. This magnitude is
fixed in the vector meson production by the reaction kinematics:
$x_p \sim (m_J^2+Q^2+|t|)/(s y)$. As a result, the relevant
$A_{ll}$ asymmetry decreases with growing energy.

Thus, we have found that the form of $A_{ll}$ asymmetry depends
on the Pomeron coupling structure. Our results show the
essential role of the " full block asymmetry"  in the $A_{ll}$
asymmetry of vector meson production. The asymmetry is vanishing at
high energies and at HERA energy the asymmetry will be
negligible. We can conclude that the spin--structure of the
Pomeron coupling  determined by  large--distance
effects in QCD can be studied in  diffractive processes.
Moreover, these reactions can give an information on the
nonforward  gluon distributions in the proton.


\begin{thebibliography}{99}

\bibitem{h1_zeus} ZEUS Collaboration,  M.Derrick et al.,  Z.Phys. {\bf C68}, 569 (1995);\\
H1 Collaboration,  T.Ahmed et al., Phys.Lett. {\bf B348},
 681 (1995).
\bibitem{rys} M.G.Ryskin, Z.Phys {\bf C57}, 89 (1993);\\
S.J.Brodsky at al., Phys.Rev., {\bf D50}, 3134 (1994).
\bibitem{mod}
A.Donnachie, P.V.Landshoff, Nucl.Phys., 1989, {\bf
B311} 509;\\
J.R.Gudell, Nucl.Phys., {\bf B336} (1990) 1.
\bibitem{lansh-m} A.\ Donnachie, P.V.\ Landshoff,
                  Nucl.\ Phys.\ {\bf B244}, 322 (1984).
\bibitem{nach} T. Arens, M. Diehl, O. Nachtmann, P. V. Landshoff
     Z.Phys. {\bf C74}, 651 (1997).
\bibitem{klen} J.Klenner, A.Sch\"afer, W.Greiner, Z.Phys {\bf A352} (1995) 203.
\bibitem{gol_mod}  S.V.Goloskokov, S.P.Kuleshov, O.V.Selyugin,
           Z.Phys. {\bf C50},  455 (1991).
\bibitem{kroll} M.\ Anselmino, P.\ Kroll, B.\ Pire, Z.Phys.
                {\bf C36}, 36 {1987}.
\bibitem{gol_kr} S.V.\ Goloskokov, P.\ Kroll, e-print: hep-ph/9807529.
\bibitem{diehl} M.\ Diehl,  Eur.Phys.J. {\bf C4} (1998) 497.
\bibitem{berger}E.L.Berger Phys.Rev.,{\bf D23} (1981) 1521.
\bibitem{mank} M.\ V\"anttinen. L.\ Mankiewicz, e-print:
hep-ph/9805338.
\bibitem{gola_ll} S.V.\ Goloskokov,Mod. Phys. Lett.,{\bf 12}, 173 {1997}; e-prints:
hep-ph 9506347; hep-ph 9509238.
\end{thebibliography}
\end{document}